\documentclass[sigconf]{acmart}

\setcopyright{acmcopyright}
\copyrightyear{2026}
\copyrightyear{2027}
\acmConference[Accepted in ICDCN 2027]{28th International Conference on Distributed Computing and Networking}{January 4--7, 2027}{Surathkal, India}

\usepackage{subcaption}
\usepackage{algorithm}
\usepackage{xcolor}
\usepackage{textcomp}
\usepackage{verbatim}
\usepackage{enumitem}
\usepackage[noend]{algpseudocode}
\usepackage{multirow}
\usepackage{setspace}

\usepackage[font=small,skip=0pt]{caption}

\setcopyright{none}
\acmDOI{}
\acmISBN{}
\renewcommand\footnotetextcopyrightpermission[1]{}
\title{Learning based Scalable Multi-Objective Scheduling of Workflows across UD-IaaS-FaaS Systems}

\author{G Vasantha Reddy}
\orcid{0000-0001-5669-9808}
\affiliation{%
  \institution{IIT Guwahati}
  \city{Guwahati}
  \country{India}
}
\email{r.vasantha@iitg.ac.in}

\author{Ram Bansal}
\authornote{These authors contributed equally to this work.}
\affiliation{%
  \institution{IIT Guwahati}
  \city{Guwahati}
  \country{India}
}
\email{ram.bansal@iitg.ac.in}

\author{Krashna }
\authornotemark[1]
\affiliation{%
  \institution{IIT Guwahati}
  \city{Guwahati}
  \country{India}
}
\email{krashna@iitg.ac.in}

\author{Suvarthi Sarkar}
\affiliation{%
  \institution{University of Western Australia India}
  \city{Mumbai}
  \country{India}}
\email{suvarthi.sarkar@india.uwa.edu.au}
\orcid{0009-0008-8676-6352}

\author{Aryabartta Sahu}
\orcid{0000-0002-5453-5022}
\affiliation{%
  \institution{IIT Guwahati}
  \city{Guwahati}
  \country{India}
}
\email{asahu@iitg.ac.in}



\begin{document}
\setstretch{1}
\makeatletter
\let\ACM@origbaselinestretch\baselinestretch
\makeatother
\begin{abstract}
Workflow scheduling in user device-centric edge–cloud computing has emerged as a challenging problem due to the need to jointly optimize execution performance, energy consumption, and monetary cost. Modern workflows consist of heterogeneous tasks, where compute-intensive components are well-suited for Infrastructure-as-a-Service (IaaS) resources, while fine-grained, event-driven tasks are more efficiently executed using the Function-as-a-Service (FaaS) model. 

We formulate this scheduling problem for application workflows over a hybrid User Device (UD)-IaaS-FaaS computing continuum as a multi-objective Mixed-Integer Linear Program (MILP). The formulation models the workflow execution mechanism by incorporating cumulative capacity constraints for IaaS resources and elastic resource provisioning for FaaS. To overcome the intractability of MILP, we propose a learning-based framework that combines graph neural networks, surrogate performance modeling, and a multi-sample REINFORCE approach. The proposed approach is workflow-aware, leveraging task pairing to improve data locality and structured task placement decisions. Agent training is organized into three stages: small-scale learning of UD-IaaS trade-offs, intermediate locality aware co-location, and large-scale handling of parallelism and aggregation bottlenecks in workflows. The proposed framework is validated through real-world experiments using the Pegasus Montage workflow on a custom UD-IaaS–FaaS testbed comprising a user device, multiple IaaS nodes, and serverless FaaS instances. Extensive experimental results demonstrate the improved trade-offs across makespan, energy consumption, and cost. Compared with the genetic algorithm and greedy scheduling approaches, the proposed approach reduces makespan by 10-17.5\%, energy consumption by 3-9\%, and cost by 4-10\%, while remaining within 3.5-5.6\% of the MILP optimum.

\end{abstract}

\maketitle
\section{Introduction}
In the domain of edge-cloud computing, workflow scheduling is a critical and emerging research problem driven by the need to balance performance, energy consumption, and cost. The hybrid edge-cloud is a well-known model that strategically combines the extensive, scalable power with local private resources of cloud services in the public domain \cite{EdgeCloudPlacement2025}. Contemporary applications are progressively built with intricate parts that are appropriate for various Infrastructure-as-a-Service (IaaS) virtual machines (VMs) that are frequently used for large, monolithic, or persistent tasks (on VMs), whereas Function-as-a-Service (FaaS) platforms are used to offload fine-grained, event-driven, or transient functions \cite{iaas_faas}. Resource optimization is possible with this hybrid IaaS-FaaS approach. For example, by running smaller jobs on FaaS and long running tasks on IaaS, the system saves more IaaS resources for computationally demanding tasks. This hybrid IaaS-FaaS approach enables optimal resource utilization. Notable platforms providing these services including Amazon EC2 (IaaS) and AWS Lambda \cite{cost_iaas_faas}.

An user device (UD) centric framework extends the hybrid model by incorporating the user’s own hardware as both a local compute resource and the central workflow orchestrator \cite{ud_edge}. In this architecture, the UD coordinates task dispatch, manages resource provisioning, and controls data flow across the distributed cloud environment. Given its central role, optimizing task offloading and scheduling decisions for the UD has become a key research challenge.

This UD centric model is distinguished from traditional cloud-only environments by a heterogeneous collaboration between untrusted public resources, trusted long term hired IaaS servers (managed by users) and a trusted private device (the UD) \cite{edge_compute, serverless, scheduling}. A complex trade-off between application completion time (makespan), user device energy consumption, and the financial costs to users of cloud services can result from using different offloading strategies, which can produce wildly disparate performance outcomes. But in user device centric model adds a number of special and noteworthy complications. Schedulers need to concern with the limited, shared CPU capacity on IaaS VMs, where multiple tasks may run concurrently, requiring cumulative resource constraint modeling rather than simple one job per node allocation. These factors collectively make the function offloading and scheduling problem in this environment significantly more complex than in standard cloud models. 
\nocite{pegasus}
The IaaS servers are provisioned for longer durations and remain under user control, making them comparatively more trustworthy. In contrast, FaaS resources are ephemeral and managed by third party service providers, and are therefore generally treated as unreliable. Consequently, temporary input-output data can be safely retained on IaaS servers to exploit data locality and reduce transfer overhead. However, such data persistence is not available on FaaS platforms. Instead, interactions with FaaS resources should follow a two-step data transfer mechanism, where data is securely staged through trusted storage before and after execution.

To develop an optimal offloading strategy, extensive research has investigated workflow scheduling in cloud and edge environments. However, most existing works focus on optimizing a single performance objective, such as minimizing makespan or minimizing energy consumption or cost. This is often not enough because users usually have to balance all three competing goals based on their own needs. For example, users might want to finish as quickly as possible for latency sensitive tasks or reduce cost and energy when they are on a limited budget. In such cases, single-objective optimization methods are inadequate, underscoring the need for multi-objective function offloading approaches.

Multi-objective optimization approaches are well-suited for decision making scenarios involving multiple conflicting objectives. A common and effective technique is to scalarize the disparate objectives (e.g., time, energy, and cost) into a single objective function using the weighted sum method. While formulating this scheduling problem as a Mixed-Integer Linear Program (MILP) provides a formal path to a true optimum, the underlying problem is known to be NP-hard. The specific complexities of this model particularly the pseudo-polynomial nature of the discrete-time formulation, the cumulative IaaS resource constraints, and FaaS resources make finding an exact solution computationally intractable for large-scale, real world applications. This highlights the urgent need for research into high performance heuristic and metaheuristic solutions tailored to this specific problem.

In this paper, we propose a multi-objective optimization approach to address the hybrid UD-IaaS-FaaS workflow offloading problem. We model the application as a Directed Acyclic Graph (DAG) \cite{taskDependency,DAG} and focus on optimizing application completion time, UD energy consumption, and user cost. Because these goals are incommensurable (measured in dollars, joules, and seconds), a normalized weighted-sum approach is used to combine them, which is essential to guaranteeing that user preferences are accurately reflected.

The primary contributions of this work are summarized as follows:
\begin{itemize}
\item We formulate workflow scheduling across user devices, hybrid IaaS and FaaS computing continuum as a linearly combined multi-objective optimization problem, modeled as a Mixed Integer Linear Program (MILP).

\item We propose a learning based framework that integrates Graph Neural Networks (GNNs), surrogate performance modeling and a multi-sample REINFORCE (MSR) based reinforcement learning approach to learn high quality placement strategies efficiently.

\item We propose a workflow-structure aware scheduling method that exploits task pairing to improve data locality and scalability across heterogeneous workflow sizes.

\item We develop a custom UD–IaaS–FaaS testbed comprising a user device, multiple IaaS nodes, and serverless FaaS instances to validate the proposed framework, and validated it using the Pegasus Montage workflow. Extensive experiments demonstrate the effectiveness of the proposed approach by achieving improved trade-offs among makespan, energy consumption, and execution cost.

\end{itemize}

\section{Literature Review}
Research in hybrid edge-cloud and serverless computing has extensively investigated trade-offs among performance, energy efficiency, and cost across multiple complementary directions. Significant research has focused on reducing energy consumption in IaaS data centers through VM consolidation and predictive scheduling techniques. Zakarya et al., \cite{BackFillMe2025} extends backfilling approach to cloud systems by utilizing idle host intervals for lower-priority VMs, thereby improving resource utilization and reducing energy usage. However, these approaches typically assume homogeneous infrastructure and stable request arrivals, limiting their applicability in heterogeneous, multi-tier systems with DAG-based dependencies.

Similarly, queueing theoretic methods, such as Anselmi et al., \cite{BalancedSplitting2025}, aim to minimize latency through balanced server grouping, achieving near-zero wait times without requiring job-size data. Despite their efficiency, these models rely on assumptions of homogeneous and static environments and do not account for task dependencies or inter-node communication, making them less suitable for hybrid computing scenarios. Hybrid orchestration frameworks, such as Sheshadri et al., \cite{HybridServerless2025}, integrates IaaS and FaaS by assigning baseline workflows to virtual machines and handles burst demand through serverless platforms, often leveraging machine learning for improved latency and cost minimization. However, these approaches are typically limited to service function chains, and lacks explicit multi-objective optimization, and do not support dynamic task placement across multiple computing tiers.

Federated learning frameworks, such as Zhao et al., in \cite{MGWFL2025}, address data and system heterogeneity through adaptive aggregation, improves model accuracy and convergence. However, they primarily focus on learning performance and do not incorporate resource scheduling or system-level optimization, leaving energy and communication trade-offs largely unaddressed. Similarly, workload placement approaches for edge-cloud systems, such as Wang et al., \cite{EdgeCloudPlacement2025}, employs optimization techniques to minimize cost and latency through joint VM placement and routing. These approaches assumes static environments and rely on offline optimization, limiting their adaptability to dynamic conditions and only partially accounting for energy efficiency.

Yang et al., in \cite{MOSEC2025} proposes multi objective RL approach for serverless offloading, balancing latency, cost, and energy. While effective for function-level scheduling, these methods do not address end-to-end workflow scheduling across heterogeneous computing resources. Wu et al. \cite{coldstartdqn} propose a Double Deep Q-Network (DDQN) based approach to mitigate cold starts and improve QoS by jointly optimizing execution time, energy consumption, and monetary cost in serverless edge computing. However, the approach is limited to scheduling independent tasks and does not consider workflow applications with task dependencies.

Overall, prior work has advanced energy-aware scheduling, hybrid orchestration, federated learning, and edge-cloud placement \cite{BackFillMe2025,BalancedSplitting2025,HybridServerless2025,MGWFL2025,EdgeCloudPlacement2025,dynworkflows,MOSEC2025,coldstartdqn}. However, most of these approaches are limited to static or single tier settings and do not jointly optimize latency, energy, and cost across hybrid infrastructures. This work addresses these gaps with MSR-GNN, a learning-based framework that jointly optimizes makespan, energy, and cost for DAG workflows across UD, IaaS, and FaaS resources.

\nocite{sarkar2025scientific,sarkar2025minimizing}
\section{System Model and Problem Statement}

\subsection{System Model}

\begin{figure}[!tb]
  \centering
  \includegraphics[width=0.95\linewidth]{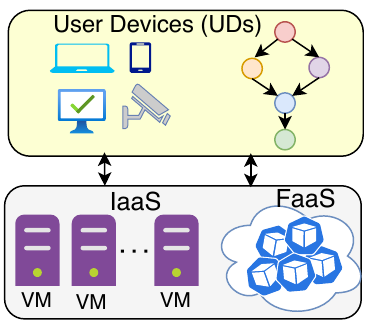}
  \caption{System Model of Hybrid UD-IaaS-FaaS Computing Continuum }
  \label{fig:architecture}
  \Description{System model diagram depicting UDs, IaaS and FaaS nodes}
\end{figure}

\autoref{fig:architecture} illustrates the proposed hybrid IaaS–FaaS system model, comprising a heterogeneous computing architecture with User Devices (UDs), Infrastructure-as-a-Service (IaaS), and Function-as-a-Service (FaaS) resources. The IaaS layer consists of multiple virtual machines (VMs) managed by a cloud service provider, whereas the FaaS layer offers containerized execution environments with heterogeneous resource configurations hosted in the cloud. User applications are modeled as Directed Acyclic Graph (DAG) workflows, where tasks are scheduled across the available UD, IaaS, and FaaS resources.

In this architecture, UDs serve as a dual role and act both as local compute resources capable of executing tasks and also, as a central orchestrator responsible for coordinating task scheduling and resource allocation across the hybrid distributed system. Upon receiving workflow tasks, the UD scheduler determines placement of tasks based on workflow requirements and system resources. Typically, the IaaS resources are utilized for compute intensive and long running tasks (as IaaS servers are hired for long term basis and managed by the users), whereas FaaS platform is suitable for lightweight and short-term executions (completely outsource as simple task execution). This hybrid design enables efficient workload distribution across tiers, thereby improving overall system performance and resource utilization.

\subsection{Workflow Model}
An application workflow is modeled as a DAG $G = (\mathcal{V}, \mathcal{E})$, 
where $V$ is the set of tasks (nodes), and $\mathcal{E} \subseteq \mathcal{V} \times \mathcal{V}$ is the set of directed edges, represents the dependencies among tasks. An edge $(t_i, t_j) \in \mathcal{E}$ denotes that task $t_j$ depends on task $t_i$. A workflow task $t_j$ is eligible for execution only after all its predecessor tasks have completed thereby satisfying its data dependencies.

The system is modeled using a discrete-time formulation, where the time horizon $H$ is partitioned into a set of equal length time slots $\mathcal{T}$, each of duration $\Delta$. This approach is often more computationally tractable for problems with complex, cumulative resource constraints compared to continuous time models. However, the choice of $\Delta$ represents a fundamental trade-off. A smaller $\Delta$ provides a more precise schedule (approaching a continuous time solution) but exponentially increases the size of the model and computational complexity, potentially making it intractable to solve.

\subsection{Constants (derived from parameters)}
These constants are derived to translate the continuous physical parameters (cycles, bytes, seconds) into the integer-based ``slot'' domain required by the discrete-time MILP formulation.

We discretize execution durations into integer slot counts. For a slot duration $\Delta$ (seconds):
\begin{equation}
D^{U}_{i} = \left\lceil \frac{c_{i}/\omega_{u}}{\Delta} \right\rceil ; 
D^{I}_{i,m,k} = \left\lceil \frac{c_{i}/f_k}{\Delta} \right\rceil ; 
D^{F}_{i,n} = \left\lceil \frac{t^{\text{cold}}_n + c_{i} / \omega_n}{\Delta} \right\rceil. \label{eq:DisTS}
\end{equation}

where $D^{U}_{i}$ denotes the execution duration of task $i$ on the user device, $D^{I}_{i,m,k}$ denotes its execution duration on IaaS node $m$ with resource tier $k$, and $D^{F}_{i,n}$ denotes its execution duration on FaaS instance $n$. The FaaS execution duration explicitly includes the cold-start delay $t^{\text{cold}}_n$, thereby accounting for the additional slot occupancy incurred during function initialization.

\subsection{Decision Variables}

Binary placement variables are defined as follows: $x^{U}_{i} \in\{0,1\}$, where $x^{U}_{i}=1$ if task $i$ is executed on the user device; $x^{I}_{i,m} \in\{0,1\}$, the value of $x^{I}_{i,m}=1$ if task executes on IaaS server $m$; and $x^{F}_{i,n} \in\{0,1\}$, the value $x^{F}_{i,n}=1$ if task executes on FaaS endpoint $n$. The scheduling (time-slot) variables are defined as follows.
\begin{itemize}
\item Start indicator: $u_{i,t}\in\{0,1\}$: 1 if the task $i$ starts at slot $t$. It is a time-indexed variable, and it creates a binary variable for each possible start time $t$ for each task $i$.
\item Start slot integer: $s_{i}\in\{0,\dots,H-1\}$ such that $s_{i}=\sum_t t\cdot u_{i,t}$.
\item Duration in slots (chosen): $D^{\text{choice}}_{i}\in\{0,\dots,H\}$ (integer).
\item Finish slot: $F_{i}\in\{0,\dots,H\}$ (integer) with $F_{i}=s_{i}+D^{\text{choice}}_{i}$.
\end{itemize}

\subsection{Constraints}
Constraints are the rules and limits that make up the ``feasible set'' of solutions. All of the following algebraic constraints must be met by any valid schedule. 

\emph{Placement exclusivity}: Each task must be placed/scheduled at exactly one location either on the UD, one IaaS server, or one FaaS endpoint, and this can be expressed as \begin{equation}\label{eq:place_excl}
x^{U}_{i} + \sum_{m\in\mathcal{S}^I} x^{I}_{i,m} + \sum_{n\in\mathcal{S}^F} x^{F}_{i,n} = 1,
\quad \forall i.
\end{equation}

\emph{IaaS tier linking}: If a task is assigned to IaaS server $m$, exactly one tier must be selected on that server:
\begin{equation}\label{eq:tier_link}
\sum_{k=1}^{\kappa} y_{i,m,k} = x^{I}_{i,m}, \quad \forall i,m.
\end{equation}
This equation links the general placement variable $x^{I}_{i,m}$ to the specific tier selection $y_{i,m,k}$. If a task is not placed on server $m$ (LHS=0), it cannot select a tier (RHS=0). If it is (LHS=1), it must select exactly one tier (RHS=1).

\emph{Start indicator consistency (one start if scheduled)}: Exactly one start slot indicator must exist if and only if the task is scheduled:
\begin{equation}\label{eq:one_start}
\sum_{t\in\mathcal{T}} u_{i,t} = x^{U}_{i} + \sum_{m,k} y_{i,m,k} + \sum_{n} x^{F}_{i,n}, \quad \forall i.
\end{equation}
Since the exclusivity constraint in \autoref{eq:place_excl} requires the RHS to equal 1, the model selects exactly one start time $t$ for each task, thereby enforcing $u_{i,t}=1$ for a single time slot.

\emph{Start slot definition}: The start time $s_i$ is determined from the binary start indicators as:
\begin{equation}\label{eq:start_def}
s_{i} = \sum_{t\in\mathcal{T}} t\cdot u_{i,t}, \quad \forall i.
\end{equation}
Since the exclusivity constraint in \eqref{eq:one_start} ensures that exactly one $u_{i,t}$ equals 1, the above expression uniquely assigns $s_i$ to the corresponding start time slot.

\emph{Execution duration selection (location-dependent)}: The execution duration is determined by the selected execution resource node:
\begin{equation}\label{eq:D_choice}
D^{\mathrm{choice}}_{i}
=x^{U}_{i}\,D^{U}_{i}+\sum_{m\in\mathcal{S}^I}\sum_{k=1}^\kappa
 y_{i,m,k}\,D^{I}_{i,m,k}+\sum_{n\in\mathcal{S}^F}
 x^{F}_{i,n}\,D^{F}_{i,n}
\quad \forall i \end{equation}
where the placement variables are binary and the execution durations $D^{\cdot}$ are constants, $D^{\mathrm{choice}}_{i}$ equals to the execution duration of the selected resource and is therefore an integer.

\emph{Finish time}: Finish slot equals start plus execution duration:
\begin{equation}\label{eq:finish_def}
F_{i} = s_{i} + D^{\text{choice}}_{i}, \quad \forall i.
\end{equation}

\emph{IaaS per-slot CPU capacity}: For each IaaS server $m$ and time slot $t$, the total CPU demand of all tasks executing during slot $t$ must not exceed the server capacity $\Omega_m$. A task assigned to tier $k$ contributes $f_k$ CPU cycles/sec over its execution interval, which is captured by summing its start time indicators across the corresponding active time slots. The resulting linear constraint is as follows:

\begin{equation}\label{eq:iaas_cap}
\sum_{i} \sum_{k=1}^{\kappa} f_k \cdot \sum_{\tau=\max(0,t-D^{I}_{i,m,k}+1)}^{t} u_{i,\tau} \cdot y_{i,m,k}
\ \le\ \Omega_m,\quad \forall m,t.
\end{equation}

This constraint is a cumulative resource constraint, formulated using a ``time-bucket'' or ``sliding window'' approach. For every server $m$ and every time slot $t$, it calculates the total CPU load. The inner sum $\sum_{\tau=\dots}^{t} u_{i,\tau}$ checks if task $i$ started at a time $\tau$ such that it is still running during slot $t$. If so, the CPU demand $f_k$ is added to the total resource load and this total load must not exceed the server's capacity $\Omega_m$. This correctly models a shared cloud resource, which is more complex than a simple IaaS model where one job occupies one node.

\emph{Precedence constraint}: This is the set of rules that makes sure both the logical dependencies of the workflow and data flow are followed:

\begin{equation}
    t^s_j \ge t^f_{j-1} + placement_{j-1,loc} \cdot \Psi^{pred}_{j-1,loc}
\end{equation}

where $t_j^s$ denotes the start time of task $t_j$, $t_{j-1}^f$ denotes the finish time of its predecessor, and $placement_{j-1,loc} \cdot \Psi^{pred}_{j-1,loc}$ represents the data transfer time required to transmit the predecessor’s output from its execution location to the current execution node.

\emph{Memory feasibility}: A workflow task can only be assigned to a resource node with sufficient available memory.

\begin{align}
x^{U}_{i}\cdot \text{mem}_{i} &\le \text{Capmem}_{u}, \quad \forall i,\\
x^{I}_{i,m}\cdot \text{mem}_{i} &\le \text{Capmem}_m,\quad \forall i,m.
\end{align}
FaaS memory feasibility is enforced by permitting task placement only on FaaS instances that satisfy the required memory capacity of the current task.

\emph{Start/finish bounds}: All $s_{i}, F_{i}, t^{\text{ud\_ready}}, t^{\text{ud\_allready}}$ are within $[0,H]$ and integer. For the workflow, we model this by introducing a continuous variable $T_{\text{app}}$ for workflow completion time and constraints $T_{\text{app}}\ge F_{i}.\Delta$ for all tasks $i$.

\subsection{Performance Models}
The performance models define how the primary objectives time, energy, and cost are computed from the solver’s decision variables. While scheduling constraints (e.g., \autoref{eq:iaas_cap}) rely on discretized slot durations $D^\cdot$, the objective functions use continuous time formulations for higher accuracy (e.g., $c_i/\omega_u$ in seconds instead of $\Delta \cdot D^{U}_{w}$).

\subsubsection{Workflow Completion Time (Makespan)}

The workflow execution time is modeled using a continuous variable $T_{\text{app}}$, constrained such that $T_{\text{app}} \geq F_i \cdot \Delta$ for all workflow tasks $i$, where execution time depends on the chosen placement. While discretized execution durations $D^{\cdot}$ are used for scheduling decisions, continuous formulations are applied for computing energy and monetary costs. The upload and data transfer time for each task $i$ is given by $T^{\text{up}}_{i} = \Delta \cdot L_i$, where $L_i$ represents the selected communication latency.
\nocite{10487075}
\subsubsection{Energy Model} Energy is modeled as conservative linear energy accounting for both local and remote resources per task $i$:

\begin{align}
E^{\text{loc}}_{i} &= x^{U}_{i}\cdot p^{\text{comp}}_{u}\cdot \frac{c_{i}}{\omega_{u}},
\end{align}

\begin{align}
E^{\text{remote}}_{i} &= (x^{I}_{i,\cdot} + x^{F}_{i,\cdot}) \cdot \Bigl( p^{\text{tx}}_{u}\cdot T^{\text{up}}_{i} + p^{\text{idle}}_{u}\cdot (T^{\text{pred}}_{i} + T^{\text{exec}}_{i}) \Bigr),
\end{align}

where $T^{\text{pred}}_{i}$ denotes the waiting time at the UD for predecessor outputs. The UD energy model considers three power states: computation ($p^{\text{comp}}_{u}$), transmission ($p^{\text{tx}}_{u}$), and idle waiting ($p^{\text{idle}}_{u}$) \cite{energy}. Consequently, local execution incurs computation energy, whereas remote execution incurs transmission and idle energy. To preserve this linearity, the remote energy terms are linearized using auxiliary variables and constraints that are activated only when the corresponding placement variable is selected.

\begin{align}
    E_{tot}=\sum_{i} E^{\text{loc}}_{i}+E^{\text{remote}}_{i})
\end{align}
where the total workflow energy consumption $E$ is computed as the sum of the energy consumed by local and remote executions. 

\subsubsection{Monetary Cost Model}

Monetary cost of each user workflow is modeled as follows: 
\begin{align}
C_{tot} = &\sum_{i\in V} \Biggl(\sum_{m\in\mathcal{S}^I}\sum_{k}
y_{i,m,k}\, c^{I}_{\text{cpu},m} \frac{c_{i}}{f_k} \nonumber + 
\sum_{m\in\mathcal{S}^I} c^{I}_{\text{vm},m}\, x^{I}_{i,m} \cdot \left( \Delta \cdot D^{\text{choice}}_{i} \right) \quad \nonumber\\
&\quad + \sum_{n\in\mathcal{S}^F} x^{F}_{i,n} \bigl( c^{F}_{\text{inv},n} + c^{F}_{\text{mem},n}\, \text{mem}_{i}\, T^{\text{exec}}_{i} \bigr) \Biggr).
\label{eq:cost}
\end{align}
where, the IaaS cost (term 2) is computed as the product of the per-second VM instance price $c^{I}_{\text{vm},m}$ and the execution duration of the task. The FaaS cost (term 3) consists of a fixed per-invocation charge $c^{F}_{\text{inv},n}$, and resource consumption cost given by the memory pricing rate $c^{F}_{\text{mem},n}$ (e.g., cost per MB-second), multiplied by the allocated memory and the precise execution time of the task \cite{optimizeCost, edgeCost}. The network transfer money costs are added by including $c_{\text{tx}}\cdot T^{\text{up}}_{i}$ terms.

\subsection{Objective Function}

\begin{equation}\label{eq:optimize}
  \min \text{TEC} = \min \left(w_T \frac{T_{\text{app}}}{T_{\text{ref}}} + w_E \frac{E_{\text{tot}}}{E_{\text{ref}}} + w_C \frac{C_{\text{tot}}}{C_{\text{ref}}}\right)
\end{equation}
where the \autoref{eq:optimize} represents the objective function formulated as a weighted combination of execution time ($T_{app}$), total energy consumption ($E_{\text{tot}}$), and total cost ($C_{\text{tot}}$). The parameters T$_{\text{ref}}$, $E_{\text{ref}}$, and $C_{\text{ref}}$ are normalization factors used to scale time, energy, and cost into a common dimension. The coefficients $w_T$, $w_E$, and $w_C$ denote the relative weights assigned to time, energy, and cost respectively and these coefficients are tunable based on user requirements, subject to the constraint ($w_T + w_E + w_C = 1$).

\section{MSR-GNN: GNN based multi-sample REINFORCE Approach}

Optimizing workflow task schedule and placement in a heterogeneous compute continuum is a high dimensional NP-hard problem, where workflow task dependencies, data-flow constraints, and resource heterogeneity must be jointly considered to minimize makespan time, energy, and cost (TEC). Although conventional RL approaches such as A3C or DQN offer lower variance, they are unsuitable for this setting due to dynamic workflow sizes and high simulation overhead. To address these challenges, a learning based GNN-driven multi-sample REINFORCE (GNN-MSR) framework is proposed as the core optimization engine, enabling stable policy learning in large combinatorial action spaces by reducing variance through multi-sample estimation. The end-to-end operational pipeline is defined as follows:
\begin{equation}
  (G, s) \xrightarrow{\pi_\theta} A \xrightarrow{f_\phi} \widehat{PEC} \xrightarrow{RL} \theta
\end{equation}
where $G$ denotes the workflow DAG and $s$ represents the global system state (e.g., number of active IaaS/FaaS nodes and optimization weights). The GNN-based policy processes $(G, s)$ to generate a task assignment $A$, while the surrogate GNN rapidly estimates the Policy Evaluation Cost (PEC) including time, energy and cost to provide immediate reward feedback.

\nocite{order,action}
The proposed framework integrates three key components as shown below.

\begin{enumerate}
\item GNN-based Policy Model ($\pi_\theta$): GNNs uses message passing mechanisms to encode the DAG topology by capturing task dependencies and system constraints, enabling effective node to resource placement decisions beyond purely local features.

\item GNN-based Surrogate Model ($f_\phi$): We use a supervised, differentiable GNN predicts the PEC including time, energy and cost for node placement, thereby replacing computationally expensive simulations with rapid and accurate estimations.

\item Policy Gradient Optimization: The multi-sample REINFORCE (MSR) approach iteratively refines the policy parameters $\theta$ using the rewards derived from the surrogate model (for improvement), guiding the agent toward good scheduling decisions within the large combinatorial action space.
\end{enumerate}

\subsection{State and Workflow Representation}
The workflow scheduling environment is formulated as a Markov Decision Process (MDP) that captures both the static workflow topology and dynamic infrastructural system conditions.

\subsubsection{State and Feature Encoding}

A global state is represented by the state vector $s = [N, n_{iaas}, w_T, w_E, w_C]$, which encodes the workflow jobs ($N$, i.e., number of application workflows), available IaaS resources ($n_{iaas}$) and the user defined optimization weights for makespan, energy, and cost, subject to $\sum w_i = 1$.

At the graph level, the workflow is modeled as a DAG $G = (V, E)$, where each task $v_i \in V$ is associated with a feature vector $x_i = [\text{type}_i, d_i, e_i, N]$ in $\mathcal{R}^6$. This feature vector uses one-hot encoding to represent the task type based on DAG nodes, for example (a) Type A for parallel entry level tasks, (b) Type B for intermediate processing and filtering and (c) Type C for fan-in tasks. Additionally with its normalized topological depth $d_i$, hardware specific assignment constraints $e_i$, and the workflow jobs size $N$.

\subsubsection{Constraint aware Action Space} 

To prevent the REINFORCE agent from selecting invalid or infeasible mappings, a constraint validation layer is implemented as an action mask $M_{i,j}$ that dynamically prunes the action space by enforcing three key rules based on infrastructure state $s$ and the workflow metadata:

\begin{enumerate}
  \item IaaS Dynamic Availability: Since only a subset of IaaS nodes may be active at a given time, the action mask excludes devices with index $j >n_{iaas}$ using the global state variable $n_{iaas}$, ensuring that task placement decisions consider only currently available resources.
  
  \item Data Sovereignty and Ingress/Egress Pinning: Certain tasks responsible for data ingestion and final aggregation are constrained to execute on user devices to avoid high-latency WAN transfers. Specifically, nodes are pinned by enforcing the action mask.
  
  \item Task Pairing for Data Locality: In the DAG workflows, certain tasks form strict producer-consumer pairs and must be co-located to avoid data transfer overhead.
\end{enumerate}

For any compute device $j \in D$ that is invalid for task $v_i$, the corresponding preference logit $z^{'}_{i,j}$ is assigned with $-\infty$, ensuring zero selection probability after Softmax normalization.

\subsection{Dual-GNN Architecture}
The proposed framework employs two complementary GNN models that leverages the relational inductive bias of the workflow DAG, enabling scalable generalization across varying workflow sizes without the need for structural retraining.

\subsubsection{GNN Policy Network ($\pi_\theta$)}
The GNN policy network encodes the workflow topology into actionable placement probabilities using two sequential Graph Convolutional Network (GCN) layers (hidden dimension = 64), enabling each task to aggregate structural information from its multi-hop neighborhood:

\begin{equation}
h_i^{(l+1)} = \sigma \sum_{j \in \mathcal{N}(i) \cup {i}} \frac{1}{\sqrt{\tilde{d}_i \tilde{d}j}} W^{(l)} h_j^{(l)} 
\end{equation}

where $h_i^{(l)}$ denotes the embedding of node $i$ at layer $l$, $W^{(l)}$ is the corresponding learnable weight matrix, $\tilde{d}_i$ is the degree of node $i$ including the self-loop, and $\sigma(\cdot)$ denotes the nonlinear activation function. To incorporate global contextual information, the final node embedding is concatenated with the global state vector, yielding $\tilde{h}_i = [h_i^{(l)} || s]$. The fused representation is then passed through a multi-layer perceptron (MLP) readout network to produce logits $z_i$, which are subsequently masked and normalized using the softmax function to obtain the placement policy $\pi\theta(A | s, G)$.

\subsubsection{Surrogate Performance Model ($f_\phi$)}

To eliminate the high latency of exploring all action space combinations, surrogate GNN is employed as fast performance estimator. This model operates on an augmented feature representation, where each node includes its assigned placement device, ($x_i^{surr}=[type_i || dev_i]$). The features are processed through GCN layers, followed by global mean to obtain a graph-level embedding. This embedding is concatenated with the global state $s$ and passed through the MLP to predict the scalar PEC. This framework samples assignments $a_i \sim \pi_{\theta}$ via a categorical distribution, with independent sampling for sequential tasks and structure aware sampling for parallel tasks to ensure load balancing and data locality. By embedding constraints within the probability space, surrogate model mitigates sparse reward issues in constrained combinatorial optimization.

\begin{algorithm}[!tb]
\caption{GNN Surrogate Training}
\label{alg:surrogate_training}
\begin{algorithmic}[1]

\Procedure{TrainSurrogate}{$G, N_{\text{samples}}, N_{\text{epochs}}$}
\State Initialize $\mathcal{D}_{\text{train}} \gets \emptyset$

\For{each workflow size $N_{\text{img}}$}
  \State Load DAG $G = (\mathcal{V}, \mathcal{E})$ corresponding to $N_{\text{img}}$
  
  \For{$\kappa = 1$ to $N_{\text{samples}}$}
    \State Generate random valid assignment $A^{(\kappa)}$
    \State Execute simulator to obtain $(T_\kappa, E_\kappa, C_\kappa)$
    \State Compute $PEC_\kappa = w_T*T_\kappa + w_E*E_\kappa + w_C*C_\kappa$
    \State $\mathcal{D}_{\text{train}} \gets \mathcal{D}_{\text{train}} \cup \{(G, A^{(\kappa)}, s, PEC_\kappa)\}$
  \EndFor
\EndFor

\State Initialize model parameters $\phi$

\For{epoch = 1 to $N_{\text{epochs}}$}
  \For{each sample $(G, A, s, PEC) \in \mathcal{D}_{\text{train}}$}
    \State $X \gets \text{BuildNodeFeatures}(G, A, s)$ \Comment{Feature build}
    \State $\widehat{PEC} \gets f_\phi(X, \mathcal{E}, s)$ \Comment{Forward pass}
    \State $\mathcal{L} \gets (PEC - \widehat{PEC})^2$ \Comment{Loss computation}    
    \State $\phi \gets \phi - \eta \nabla_\phi \mathcal{L}$ \Comment{Parameter update}
  \EndFor
\EndFor
\State \Return trained surrogate model $f_\phi$
\EndProcedure

\end{algorithmic}
\end{algorithm}

Algorithm \ref{alg:surrogate_training} describes the offline supervised training of the GNN surrogate model, consisting of data generation and model training phases. The process begins by initializing an null assignments to $D_{train}$ (Line 2). During assignment generation (Lines 4-9), DAGs of varying sizes are processed by sampling valid placements ($A^{(\kappa)}$), evaluating them with a simulator to obtain execution time, energy, and cost, aggregating these into $PEC_\kappa$, and storing the resulting samples in training model $D_{train}$ (Lines 6-9). After initializing model parameters, the training phase (Line 10) iterates over epochs, where each sample $(G, A, s, PEC)$ is encoded with augmented node features, passed through the GNN to predict PEC, and optimized using MSE loss $\mathcal{L} \gets (PEC - \widehat{PEC})^2$ with gradient based updates $\phi$ (Lines 13-16). Finally, the algorithm outputs a trained surrogate model capable of faster cost estimation (Line 17).

\subsection{Optimization via Multi-Sample REINFORCE}
The core objective of the solution is to minimize the multi-objective policy evaluation cost over a large combinatorial action space. To mitigate the high gradient variance inherent in standard REINFORCE, a multi-sample estimation strategy inspired by Group Relative Policy Optimization (GRPO) is employed.

\subsubsection{Group Relative Advantage Scaling}

For a given state $s$ and DAG $G$, the policy generates $\kappa$ independent assignment trajectories ${A(1),\ldots,A (\kappa)}$. Each assignment is evaluated by the surrogate model to obtain a PEC, defining the reward as $r_\kappa= -\widehat{PEC_\kappa}$. Instead of using a fixed baseline, we compute the advantage score $Adv_\kappa$ relative to the sample mean:
\begin{equation}
\label{eq:Advantage}
Adv_\kappa = r_\kappa - \frac{1}{\kappa} \sum_{j=1}^{\kappa} r_j
\end{equation}
This group relative formulation provides positive reinforcement only for good or above average assignments, improving stability and eliminating scale dependent bias across varying workflow complexities.


\begin{algorithm}[!t]
\caption{GNN-Based Constraint-Aware MSR}
\label{alg:rl_training}
\begin{algorithmic}[1]

\Procedure{TrainPolicy}{$G, f_\phi, \theta$}

\State Initialize baseline $b \gets 0$
\State Initialize replay buffers $\mathcal{S}_A, \mathcal{S}_L, \mathcal{S}_R, \mathcal{S}_H$

\For{episode $t = 1$ to $T$}

  \State Sample state $s = [N_{\text{img}}, n_{\text{iaas}}, w_T, w_E, w_C]$
  \State Load DAG $G = (\mathcal{V}, \mathcal{E})$ for $N_{\text{img}}$
  \State Build initial node features $X$
  \State $(\mathcal{S}_A, \mathcal{S}_L, \mathcal{S}_R, \mathcal{S}_H) \gets$
  \Call{GenMaskedTraj}{$G, s, X, \theta$} \Comment{Generate $\kappa$ trajectories}
  \State $\bar{r} \gets \frac{1}{\kappa} \sum_\kappa \mathcal{S}_R[\kappa]$ \Comment{Reward processing}
  \State $\tilde{r} \gets \text{Normalize}(\bar{r})$; 
  $Adv \gets \tilde{r} - b$
  \For{each trajectory $k$}\Comment{Policy gradient update}
    \State $\mathcal{L}_{\text{policy}} \gets -Adv \cdot \mathcal{S}_L[\kappa]$; 
    $\mathcal{L}_{\text{entropy}} \gets -\beta \cdot \mathcal{S}_H[\kappa]$; 
    \State $\mathcal{L} \gets \mathcal{L}_{\text{policy}} + \mathcal{L}_{\text{entropy}}$; 
     $\theta \gets \theta - \eta \nabla_\theta \mathcal{L}$
  \EndFor
  \State $b \gets \alpha b + (1 - \alpha)\tilde{r}$ \Comment{Baseline update}

\EndFor
\State \Return trained policy $\pi_\theta$
\EndProcedure

\end{algorithmic}
\end{algorithm}

Algorithm \ref{alg:rl_training} outlines the training of proposed MSR-GNN optimization of baseline approach. The procedure begins with initialization (Lines 2-3), where a running baseline $b$ is set and buffers are prepared to store trajectory data. At each episode (Lines 5-10), the state $s$ and workflow DAGs $G$ are sampled to construct node features, and $\kappa$ constraint-aware assignment trajectories are generated using the current policy. The rewards are then averaged ($\bar{r} \gets \frac{1}{\kappa} \sum_\kappa \mathcal{S}_R[\kappa]$) and normalized (${Normalize}(\bar{r})$) (Lines 9-10), and compared with the baseline $b$ to compute the advantage ($Adv \gets \tilde{r} - b$). In the update phase (Lines 12-13), each trajectory contributes to the total loss $\mathcal{L}$, combining advantage weighted policy gradients with entropy regularization, and the GNN parameters $\theta$ are updated via gradient descent $\eta \nabla_\theta \mathcal{L}$. Finally, the baseline $b$ is updated using an exponential moving average of recent rewards (Line 14) to stabilize subsequent training.

\subsubsection{Policy Gradient Update}
The policy parameter $\theta$ is updated via gradient descent to maximize the expected relative advantage, with an entropy regularization term $\mathcal{H}(\pi_\theta)$ (coefficient $\beta = 0.2$) included to encourage exploration and prevent premature convergence. 
\begin{equation}
\nabla_\theta J(\theta) \approx \frac{1}{\kappa} \sum_{\kappa=1}^\kappa \left[ Adv_\kappa \cdot \nabla_\theta \log \pi_\theta(A^{(\kappa)} | s, G) \right] + \beta \nabla_\theta \mathcal{H}(\pi_\theta)
\end{equation}
\subsubsection{Curriculum Learning Schedule}
Random initialization in large DAGs rarely yields in probability of finding efficient configurations. So, the environment state $s$ is sampled using a curriculum learning strategy. Training begins with small workflows to learn basic UD-IaaS trade-offs, progresses to intermediate sizes to capture data locality patterns, and finally scales to larger workflows. The shared message-passing weights of GNN enables effective transfer of these learned structural patterns to larger, unseen DAGs.

A predefined training schedule prevents the agent from being overwhelmed by high-dimensional action spaces before learning fundamental resource trade-offs from smaller cases. Accordingly, training is organized into three stages, each targeting key structural patterns of the scientific pipeline.

\begin{enumerate}
\item Stage 1: Structural Foundations: The agent learns basic UD-IaaS trade-offs, with small graphs enabling broad exploration and initial understanding of task classes.

\item Stage 2: Intermediate Locality: With increased overlap, the agent learns data locality by co-locating tasks and capturing fan-in and fan-out dependencies.

\item Stage 3: Production Scale: The agent manages high parallelism and the aggregation bottlenecks by generalizing learned patterns to larger workflow sizes.
\end{enumerate}

\subsection{Real-Time Inference}
Upon completion of the curriculum guided training, the MSR-GNN framework is deployed for live scheduling. During inference, the probabilistic sampling is replaced with a deterministic $\arg \max$ selection across the masked distribution:
\begin{equation}
A^* = \arg \max_{a_i \in D} \pi_\theta(a_i \mid s, G)
\end{equation}
Algorithm \ref{alg:rl_inference} describes the deterministic, low-latency inference phase of the MSR-GNN framework, executed after policy parameter $\theta$ and it is optimized via curriculum based training. For a given workflow $G$, node features ($\mathcal{H}^{\kappa}$) are processed through GCN layers to capture multi-hop dependencies, combined with the global state $s$, and passed through the MLP to produce placement logits $Z$ (Lines 3-4). A constraint mask ensures only valid assignments are considered (Line 5), and Softmax followed by $\arg\max$ yields the final placement $A^{*}$ (Line 6). Unlike the training phase, the inference procedure is a deterministic, low-latency operation suitable for real-time scheduling in dynamic IaaS-FaaS environments. This single forward pass inference operates in $O(|\mathcal{E}|)$ time enables near-optimal scheduling while decoupling training complexity from deployment.


\begin{algorithm}[!tb]
\caption{GNN-MSR Inference}
\label{alg:rl_inference}
\begin{algorithmic}[1]

\Procedure{Inference}{$G, s, \theta$}
\State Build node features $X$
\State $\mathcal{H}^{(\kappa)} \gets \text{GCNConv}(X, \mathcal{E}, \theta)$
\State $Z \gets \text{MLP}([\mathcal{H}^{(\kappa)} \parallel s])$
\State Apply constraint masking: $Z' \gets Z + M$
\State $\pi \gets \text{Softmax}(Z')$; $A^* \gets \arg\max \pi$
\State \Return optimal assignment $A^*$

\EndProcedure
\end{algorithmic}
\end{algorithm}

\section{Experimental Setup and Results}

\begin{figure}[tb!]
  \centering
  \includegraphics[width=1\linewidth]{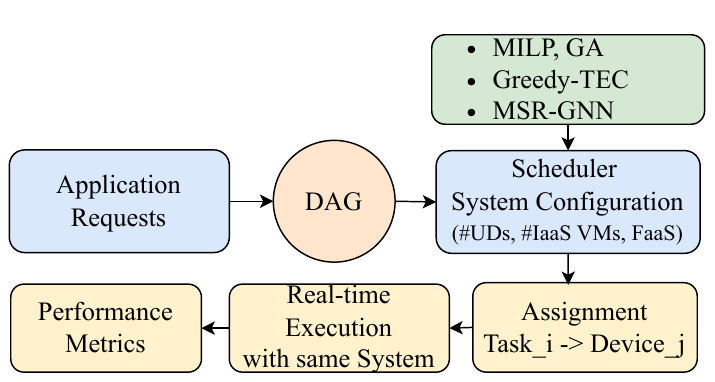}
  \caption{Workflow Deployment and Execution flow}
  \label{fig:deployment}
  \Description{This image shows real-time deployment of proposed solution and its execution flow}
\end{figure}

\begin{figure*}[tb!]
  \centering
  \includegraphics[width=0.97\linewidth]{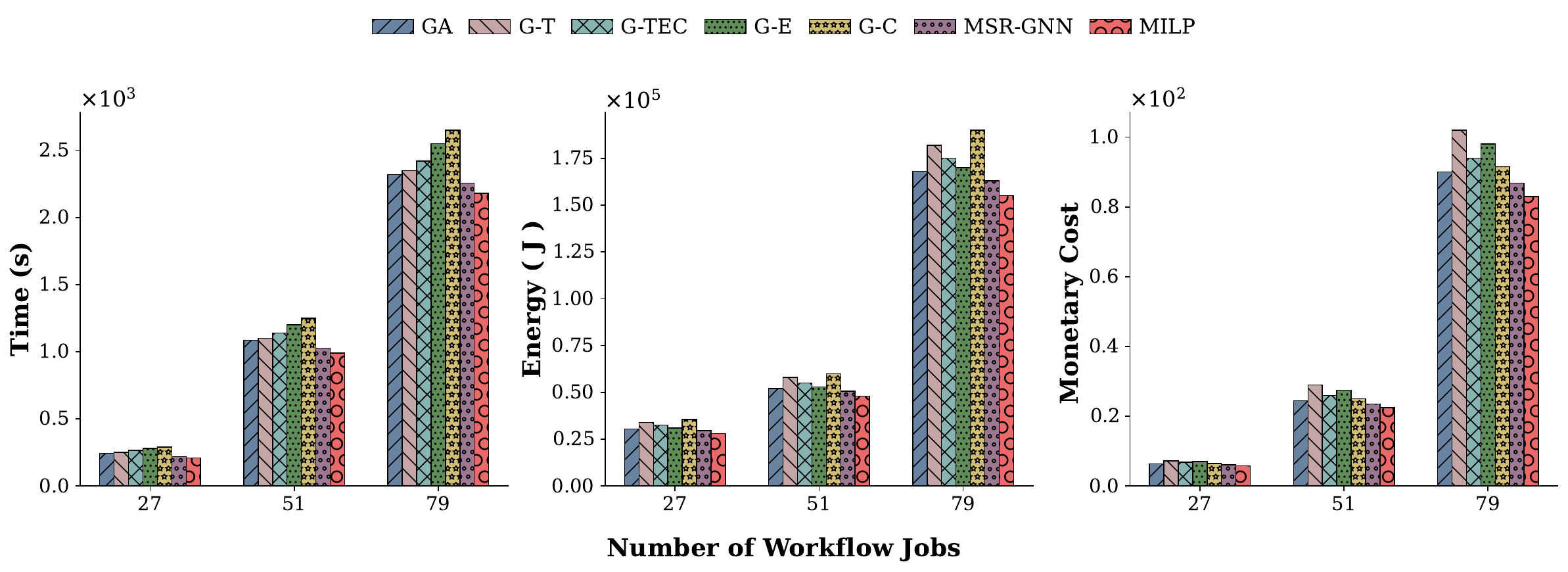}
  \caption{ (a) Execution Time, (b) Energy Consumption and (c) Monetary Cost under varying number of workflow jobs}
  \Description{this figure shows Execution Time (left), Energy (middle) and Monetary Cost (right) performance under varying number of Jobs}
  \label{fig:tec}
\end{figure*}

\begin{table}[tb!]
\centering
\caption{Experimental Parameters Used in the Evaluation}
\label{tab:exp_params}
\renewcommand{\arraystretch}{1.2} 
\begin{tabular}{@{}llc@{}}
\toprule
\textbf{Parameter Description} &  \textbf{ Value Range} \\
\midrule
Workflow Job Size &  2 - 80 \\
Makespan & 0.5 - 2500 Sec \\
Energy Consumption & 5000 - 150000 Joules \\
Monetary Cost & 1 - 100 units \\
Normalized Weights ($W_T, W_E, W_C$) & 0.005 - 0.99 \\
CPU Requirement of Jobs &  0.5 - 3 vCPU\\
Memory Requirement of Jobs & 128 - 500 MB\\
\bottomrule
\end{tabular}
\end{table}

To evaluate the proposed MSR-GNN framework, we develop a comprehensive heterogeneous UD-IaaS-FaaS testbed that emulates a realistic compute continuum integrating IaaS VMs and FaaS resources. While serverless capabilities are implemented through a local FaaS deployment using Kubernetes cluster nodes with Knative serverless framework. The testbed comprises a centralized controller for workflow orchestration and scheduling, together with distributed compute nodes spanning both resource tiers. Real-world workflows, such as Pegasus Montage image processing workflow \cite{montage}, face detection, log processing, and image processing applications are deployed to generate realistic execution traces (Ref. \autoref{tab:exp_params}) and validate the proposed framework. The energy consumption on the UD (laptop) is measured by PowerTop and EnergyTop profiler of Linux system. The system architecture and hardware configuration of the testbed are summarized in \autoref{tab:system_config}.

\begin{table}[tb!]
\caption{System Configuration}
\label{tab:system_config}
\begin{tabular}{|l|c|l|}
\hline
Components & Count & Configuration \\ \hline
Controller & 1 & CPU: Quad-core, RAM: 16 GB\\
& & OS: Ubuntu24.0 \\ \hline
IaaS Nodes & 5 & CPU: 4-8 cores, RAM: 8-16 GB\\
& & OS: Ubuntu22.04 \\ \hline
FaaS Nodes & 3 & Platform: Kubernetes\\
& & Framework: Knative \\ \hline
\end{tabular}
\end{table}

\nocite{cpcp,dataparallel}

\subsection{Deployment and Execution}

\autoref{fig:deployment} illustrates the execution of a typical workflow and the corresponding task-to-resource assignments on the testbed. The final task assignment $A^{*}$ is serialized into a \texttt{NODE-ASSIGNMENT} dictionary, mapping task IDs to resource labels (\texttt{laptop}, \texttt{iaas\_1}, \texttt{faas}). This mapping is then provided to the workflow executor, which manages container orchestration and data transfers across the heterogeneous compute continuum.

\subsection{Compared Approaches}
\begin{itemize}
\item The MILP formulation jointly determines task placement (UD/IaaS/FaaS), execution tier, and start times using binary decision variables and linear constraints, providing a lower bound on TEC due to its exhaustive search, even though its practical deployment is not considered.
\item Greedy methods include four variants: (a) Greedy-TEC (minimizes instantaneous TEC), (b) Greedy-T (execution time or makespan), (c) Greedy-E (energy), and (d) Greedy-C (cost), each making local decisions based on a single objective.
\item The Hybrid Genetic Algorithm (GA) initializes its population using greedy solutions to ensure diversity and faster convergence, with fitness evaluated via a simulation model equivalent to the MILP TEC formulation, enabling consistent comparison across all approaches.
\end{itemize}

\subsection{Execution Time (Makespan) Analysis}

\autoref{fig:tec}(a) shows the makespan results of the proposed MSR-GNN framework across different workload scales (27, 51 and 79 workflow jobs). Compared to baseline approaches, the proposed MSR-GNN approach achieves significant improvements, reducing execution time up to 10.7\% relative to GA and up to 17.5\% compared to Greedy-TEC. Moreover, the framework consistently attains solutions within 3.5 to 4.1\% of the MILP solution. These results highlight the GNN-MSR capability optimize makespan with near-optimal accuracy, while avoiding the exponential computational complexity of offline MILP method.

\subsection{Energy Consumption Analysis}

\autoref{fig:tec}(b) illustrates the energy consumption trends across the varying DAG workflow sizes (27, 51 and 79 workflow jobs). The experimental results indicate that the MSR-GNN framework consistently reduces energy consumption across all the scenarios. Compared to baseline methods, MSR-GNN achieves overall energy savings of up to 3.1\% over GA and up to 9.1\% over Greedy-TEC. In comparison with MILP, the framework demonstrates strong performance, producing energy-efficient placements within a margin of 5.1 to 5.5\% of the MILP solution.

\subsection{Cost Analysis}

\autoref{fig:tec}(c) shows the monetary cost results across different approaches, the proposed MSR-GNN framework consistently outperforms baseline approaches in minimizing cost across varying size of application workflows (27, 51 and 79 workflow jobs). As compared to the traditional GA and Greedy-TEC approach, proposed MSR-GNN approach achieves cost reductions up to 4.3\% and 9.9\%, respectively. The MSR-GNN approach effectively navigates the high dimensional scheduling space, producing near-optimal placements within 4.5-5.6\% of the MILP across all workflow scales.

\begin{figure}[tb!]
\centering
\includegraphics[width=0.47\textwidth]{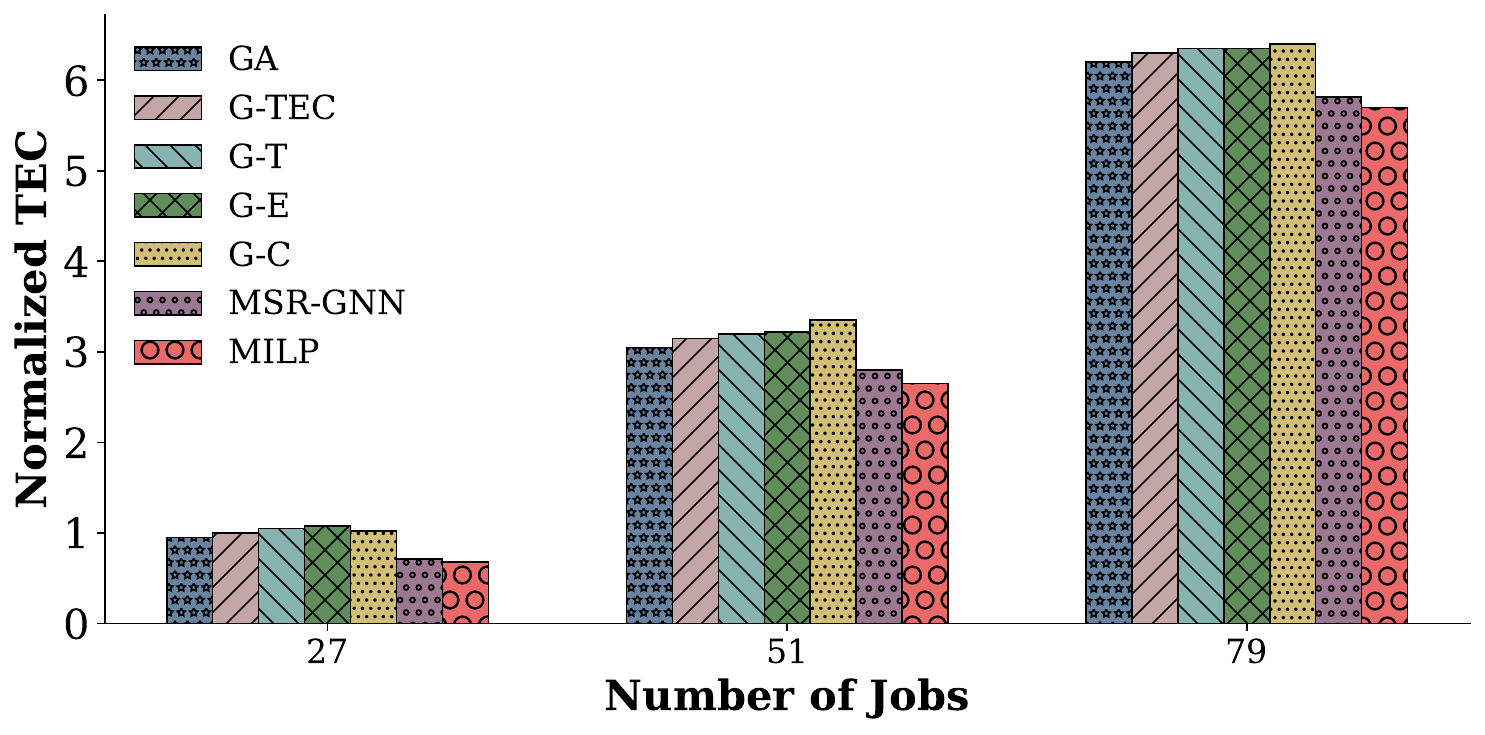}
\caption{Normalized TEC across varying workflow sizes}
\Description{This figure shows the results comparison between normalized TEC vs varying number of workloads}
\label{fig:jobs_tec}
\end{figure}

\begin{figure}[tb!]
\centering
\includegraphics[width=0.46\textwidth]{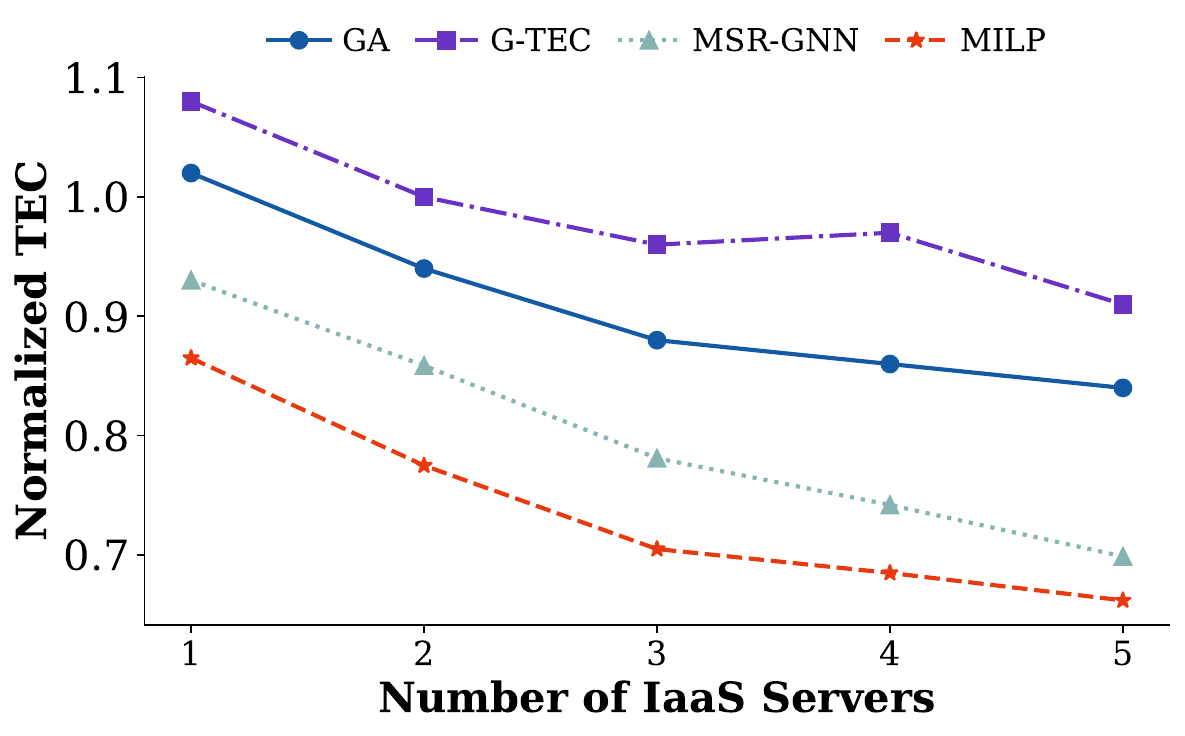}
\caption{Normalized TEC vs. Varying number of IaaS nodes}
\Description{This figure shows Normalized TEC performance against varying number of IaaS Nodes (VMs) }
\label{fig:iaas_tec}
\end{figure}

\subsection{Normalized TEC Analysis}
The \autoref{fig:jobs_tec} shows the normalized execution time (makespan), energy, and cost (TEC) across different workflow scales (27 nodes, 51 nodes, and 79 nodes), demonstrating the effectiveness of the proposed MSR-GNN framework. Compared to baseline methods, MSR-GNN achieves significant improvements, reducing TEC by up to 25.0\% relative to GA and up to 28.8\% compared to Greedy-TEC. Also, our approach consistently attains solutions within 2.0-5.7\% of the MILP optimum. These results highlight the GNN-MSR capability to jointly optimize makespan, energy, and cost with near-optimal accuracy, while avoiding the exponential computational complexity of offline optimization methods.

\subsection{Impact of IaaS Resource Availability}

\autoref{fig:iaas_tec} shows normalized TEC values for varying numbers of IaaS nodes (1 to 5) to evaluate adaptability under different infrastructure capacities. As resources increase, the proposed MSR-GNN framework consistently outperforms baseline approaches, achieving normalized TEC reductions of up to 16.8\% compared to the GA and up to 23.5\% against the G-TEC approach. the MSR-GNN agent maintains highly competitive performance, yielding placement strategies that fall within a narrow margin of just 5.5 to 10.8\% of absolute optimality of MILP.

\subsection{Impact of Objective Weight Variation}

\begin{figure}[tb!]
\centering
\includegraphics[width=0.47\textwidth]{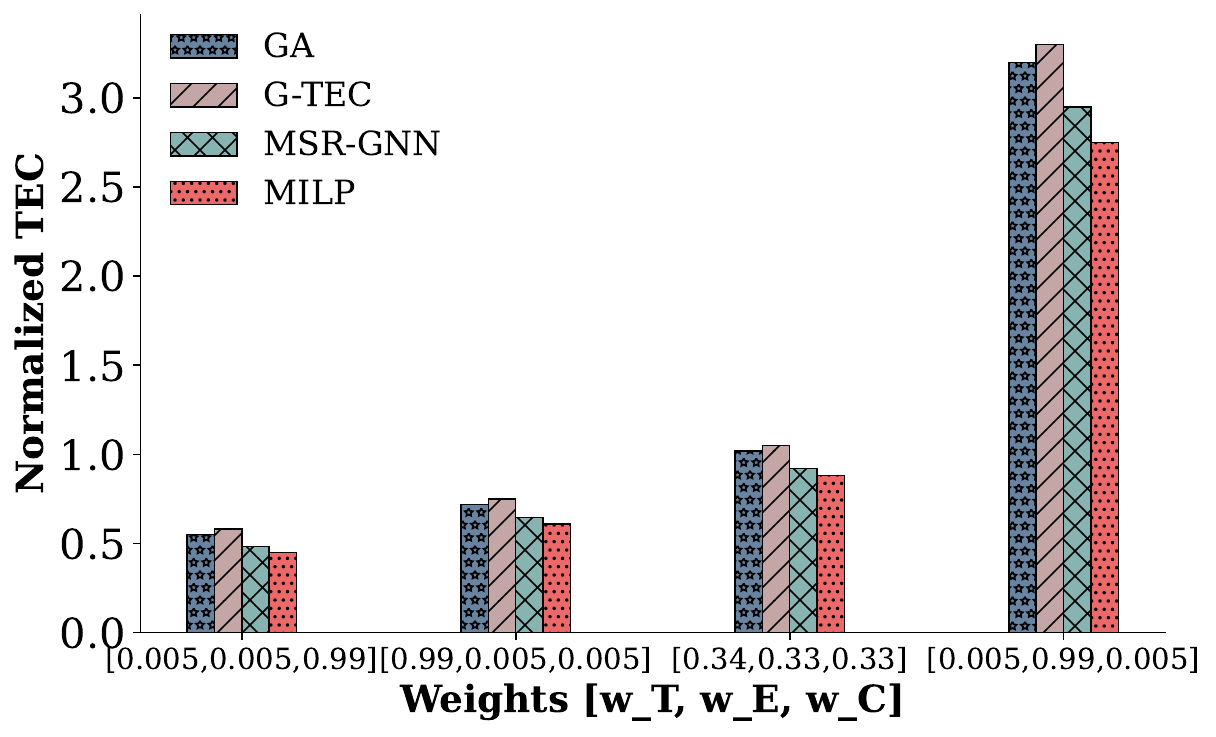}
\caption{Normalized TEC under varying objective weights}
\label{fig:weight_tec}
\Description{This figure shows Normalized TEC performance against varying Objects weights for Time, Energy and cost respectively }
\end{figure}

\autoref{fig:weight_tec} shows normalized TEC values for different weight configurations ($[w_T, w_E, w_C]$): time-focused $[0.99, 0.005, 0.005]$, balanced $[0.34, 0.33, 0.33]$, energy-focused $[0.005, 0.99, 0.005]$, and cost-focused $[0.005, 0.005, 0.99]$. The results highlight significant variation in TEC due to inherent trade-offs among objectives, where optimizing one metric may adversely affect others. Despite this, the proposed MSR-GNN framework consistently outperforms baseline heuristics, achieving TEC reductions of 7.8-12.4\% over GA and 10.6-16.9\% over Greedy-TEC. Furthermore, it maintains close proximity to the MILP optimum, with a narrow gap of 4.5-7.3\% across all scenarios, demonstrating near-optimal performance under diverse objective preferences.

\begin{figure}[tb!]
\includegraphics[width=0.47\textwidth]{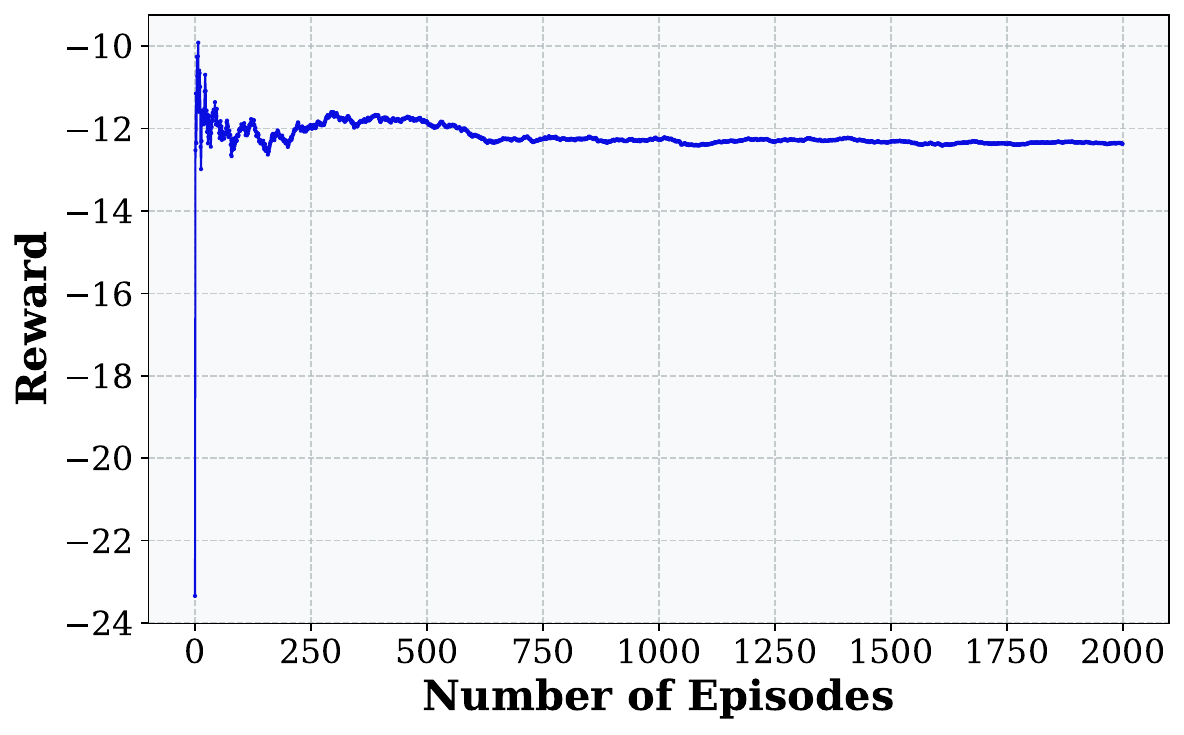}
\caption{Learning curve convergence: Reward vs. No. of Episodes}
\Description{This figure shows learning curve of proposed method's training, X-axis: number of Episode Y-axis: Rewards }
\label{fig:learning_curve}
\end{figure}

\subsection{Learning Behavior of the MSR-GNN Policy}

To analyze the convergence of proposed MSR-GNN framework, training reward is monitored across all the training episodes, as shown in \autoref{fig:learning_curve}. A high variance is observed during the initial phase (0 to 500 episodes) due to stochastic action space exploration. however, the RL agent gradually stabilizes by leveraging surrogate cost feedback to reinforce efficient placements. The learning curve flattens beyond 1000 episodes, which indicates the convergence to a near-optimal policy. After approximately 1,250 episodes, the reward exhibits negligible variance, suggesting diminishing returns from further training and the risk of overfitting, which may reduce generalization to unseen workflows. Therefore, early stopping within 800 to 1100 episodes provides an effective balance between convergence and generalization.

\section{Conclusion and Future Work}

This work proposes a scalable and adaptive workflow scheduling framework for heterogeneous IaaS-FaaS systems using a GNN based multi-sample REINFORCE approach. The problem is formulated as a MDP, where GNNs capture task dependencies and global system dynamics, and a differentiable surrogate enables faster PEC estimation, reducing reliance on expensive simulations. Experimental results show that proposed MSR-GNN outperforms heuristic baseline approaches, achieving near-optimal performance within 3.5-5.6\% of MILP, and adapts effectively to varying workflow sizes, resource availability, and objective weights. Overall, the framework offers an efficient, low-latency solution for large-scale DAG scheduling in hybrid systems.

In Future work we plan to extend this framework in several directions. First, incorporating real-time system feedback and online learning would enable adaptation to dynamic workload arrivals and fluctuating resource and network conditions. Second, we extends this framework to support multi-user and multi-tenant scenarios with fairness constraints could improve applicability in shared cloud environments.

\bibliographystyle{unsrt}
\bibliography{reference}
\end{document}